\documentclass[pra,a4paper,groupedaddress,showkeys,showpacs]{revtex4}
\usepackage{graphicx}
\usepackage{tabularx}
\usepackage{amsmath}
\usepackage{amsfonts}
\usepackage{amssymb}
\usepackage{natbib}

\begin{document}

\title{All-optical atom surface traps implemented with one-dimensional planar diffractive microstructures}
\date{\today}
\author{O. Alloschery}\author{R. Mathevet}\author{J.
Weiner}\email{jweiner@irsamc.ups-tlse.fr} \affiliation{IRSAMC/LCAR\\
Universit\'e Paul Sabatier, 118 route de Narbonne,\\31062
Toulouse, France}


\begin{abstract}
We characterize the loading, containment and optical properties of
all-optical atom traps implemented by diffractive focusing with
one-dimensional (1D) microstructures milled on gold films. These
on-chip Fresnel lenses with focal lengths of the order of a few
hundred microns produce optical-gradient-dipole traps. Cold atoms
are loaded from a mirror magneto-optical trap (MMOT) centered a
few hundred microns above the gold mirror surface. Details of
loading optimization are reported and perspectives for future
development of these structures are discussed.
\end{abstract}


\maketitle

\section{Introduction}

Manipulation and control of matter at the micro- nano- and atomic
level has become increasingly important for the investigation of
cold quantum gases\,\cite{HHH01}, atom
interferometry\,\cite{WAB05}, quantum information
processing\,\cite{WHS06} and precision positioning of atoms on or
under surfaces\,\cite{MHP04,LRH05}.  For the most part magnetic
forces have been used in order to take advantage of favorable
scaling laws and on-chip integration technologies as component
size reduces to the micron scale\,\cite{WL95,RHH99,CCF00}.  The
implementation of optical forces has proceeded more slowly because
the subwavelength electro-magnetic field localisation required to
achieve comparable scaling-law advantages has been more difficult
to realize. Earlier work has emphasized remote, table-top
preparation of optical trap arrays with subsequent projection into
the closed vacuum system containing the cold atoms\,\cite{DVM02}
or the use of ``optical tweezers" either for long-range transport
of cold-atom condensates\,\cite{GCL02} or for the sorting of
individual atoms in 1D strings\,\cite{MAD06}. Recent developments
in integrated- and nano-photonics\,\cite{nano}, however have
renewed interest in all-optical approaches to atom and molecule
manipulation using on-chip planar architectures\,\cite{ETP05,O06}.
Optical forces are of interest because of their wide applicability
to atoms, molecules and clusters.  They rely on electrical
polarization, independent of net magnetic dipole moment, while
most molecules are singlets with no net magnetic dipole in the
electronic ground state.  Here we report a study using planar
diffractive focusing to load an elongated optical dipole trap with
cold atoms from a 3-D mirror magneto-optical trap\,\cite{RHH99}
oriented $100-500~\mu$m above a gold mirror surface.
\section{Overall Experimental setup}
\begin{figure}
\includegraphics[width=0.75\columnwidth]{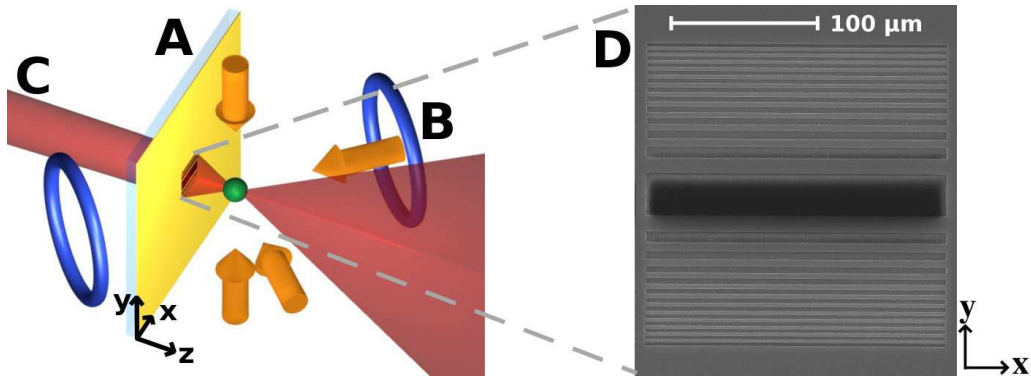}\caption{Schematic of the MMOT/FFORT setup.
A--Gold mirror focused-ion-beam (FIB)-milled at center with 1-D
Fresnel diffraction lens. B--Laser beams  and magnet coils (blue
rings) forming the MMOT. Green cloud of cold Cs atoms is trapped
at the MMOT center at a distance typically $100-500\,\mu$m from
the mirror surface. C--Fresnel far off-resonant trap (FFORT) laser
beam illuminating the mirror and Fresnel structure from the rear.
D--Zoom of the Fresnel structure at the mirror center with
500~$\mu$m focal length. Overall Fresnel motif dimensions:
209~$\mu$m $\times$ 206~$\mu$m; central slit width: $\simeq
28~\mu$m with 12 slits on each side. FFORT laser beam: focused
gaussian TEM$_{00}$ on the Fresnel structure with 230~$\mu$m
intensity ($1/e^2$) spot diameter.}\label{Fig:manipe_schema}
\end{figure}

Figure \ref{Fig:manipe_schema} shows a schematic of the overall
setup.
The MMOT, cooling and repumping near the
Cs~[$\mathrm{S}_{1/2},\mathrm{F}$=4]\,$\rightarrow$\,
Cs~[$\mathrm{P}_{3/2},\mathrm{F^{\prime}}$=5] and the
Cs~[$\mathrm{S}_{1/2},\mathrm{F}$=3]\,$\rightarrow$\,
Cs~[$\mathrm{P}_{3/2},\mathrm{F^{\prime}}$=3] transitions
respectively, produces a cloud of cold cesium atoms several
hundred microns from the gold mirror surface. The MMOT is loaded
from background Cs vapor and captures $\simeq 3\times 10^6$ atoms
with a density of $\simeq 3\times 10^{11}$ cm$^{-3}$ at a
temperature of $\simeq 30~\mu$K. The surface itself is formed from
a gold layer 400~nm thick evaporated onto a 25~mm square, 1~mm
thick fused silica substrate. Two laser beams reflect in the
horizontal $x-z$ plane while two counterpropagating beams along
the vertical $y$ axis plus the external magnetic field gradient
complete the standard six-beam MOT configuration. A focused ion
beam (FIB) is used to mill at the mirror center a pattern of
horizontal slits so as to produce a Fresnel diffraction lens with
a 1-D focus. When illuminated from behind the mirror with an
intense laser beam tuned far to the red of the Cs D2 line, the
structure forms a Fresnel far-off resonance dipole trap (FFORT).
This FFORT is loaded with cold atoms from the MMOT. We have
studied cold Cs atom trapping using diffractive structures with
focal lengths of 500~$\mu$m and 200~$\mu$m.
\subsection{Fresnel Lens Characterization}
\begin{figure}
\begin{minipage}[t]{0.45\columnwidth}\centering
\includegraphics[width=0.9\columnwidth]{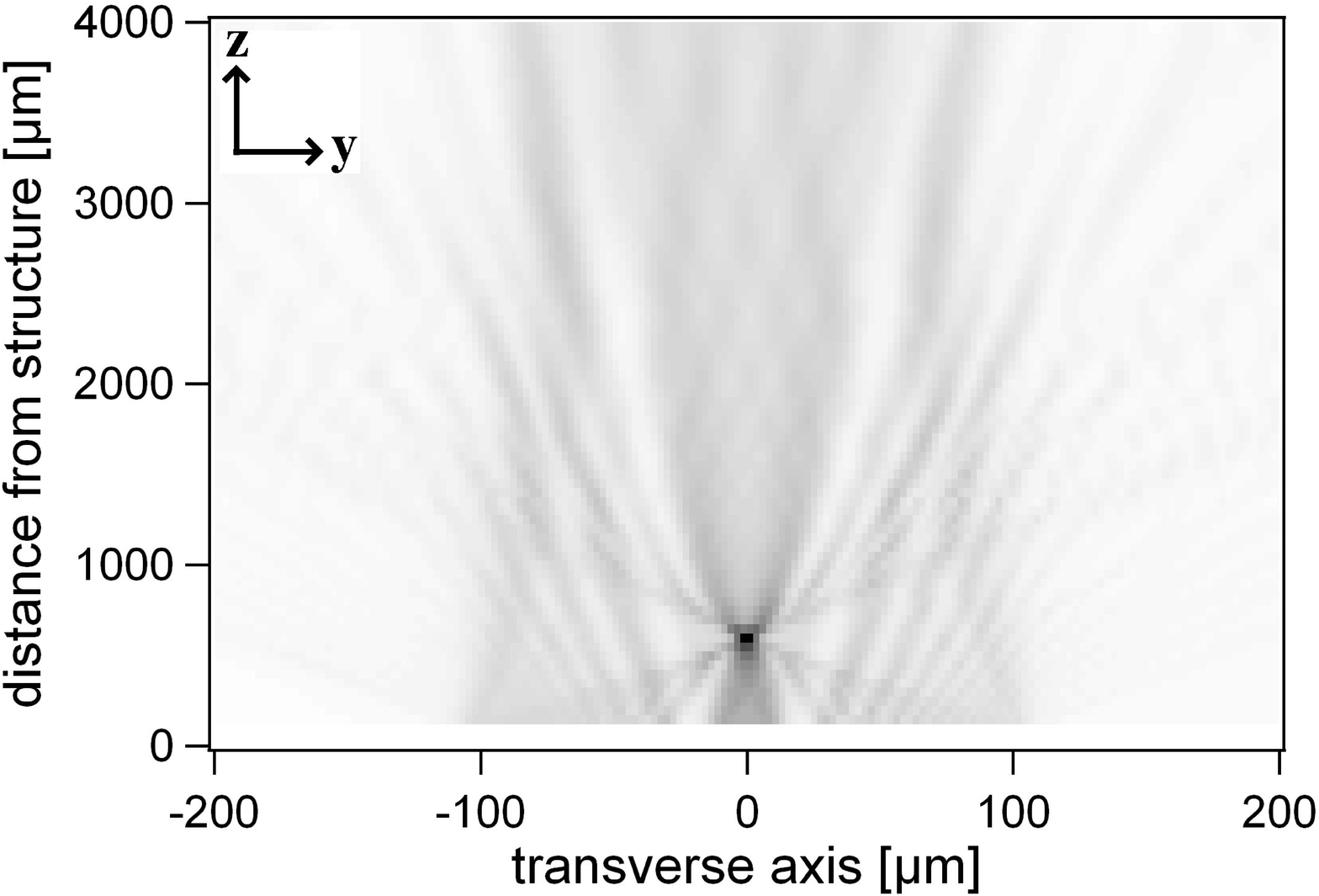}\caption{Measured spatial intensity map
in the $y-z$ plane. Coordinate axes indicated in
Fig.\,\ref{Fig:manipe_schema} and measurement setup in
Fig.\,\ref{Fig:imaging schematic}.  Focal distance is designed for
500~$\mu$m.}\label{Fig:f500experience}
\end{minipage}\hspace{0.04\columnwidth}
\begin{minipage}
[t]{0.45\columnwidth}\centering
\includegraphics[width=0.9\columnwidth]{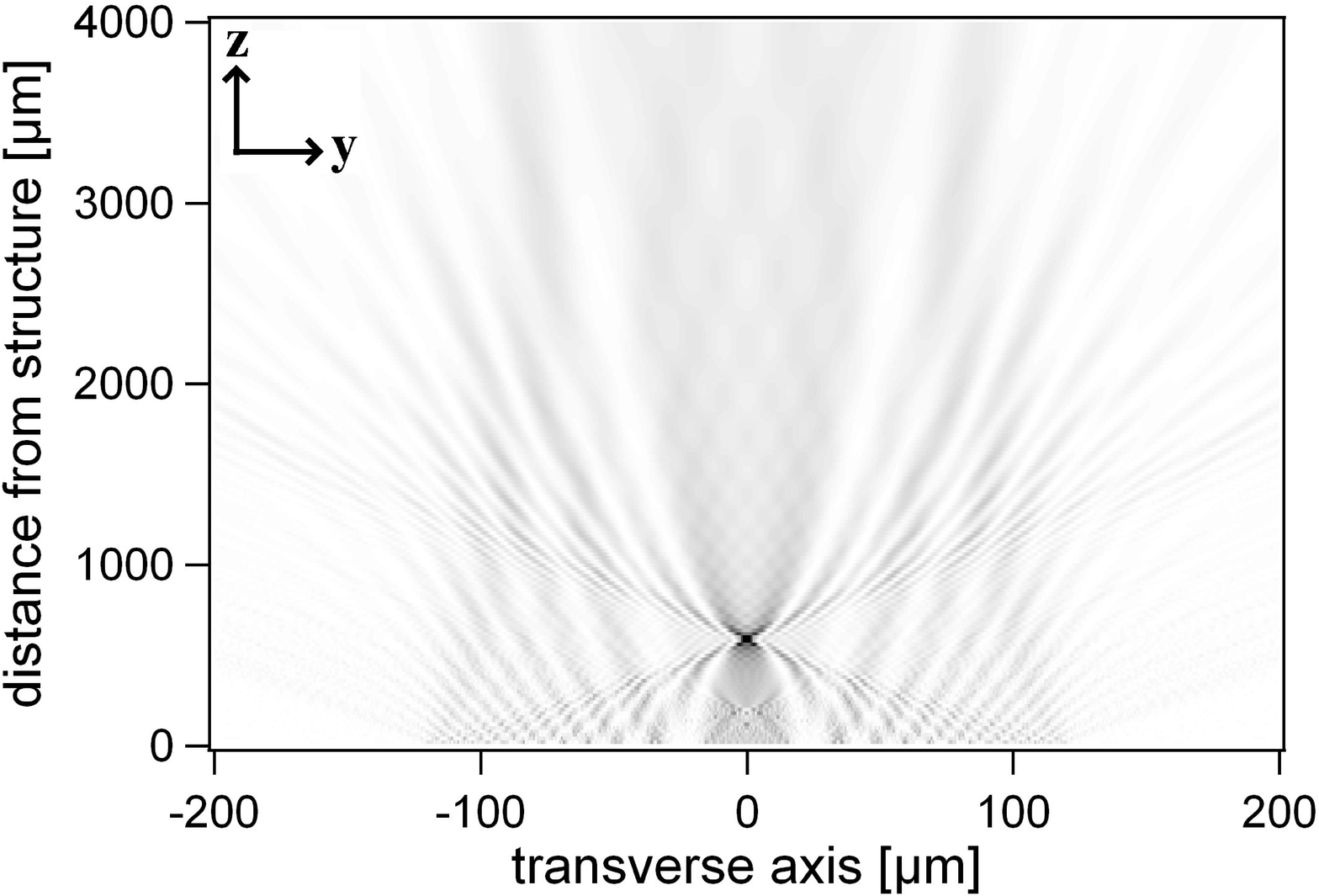}\caption{Numerical simulation of the diffractive pattern for the structure measured
in Fig.\,\ref{Fig:f500experience}.}\label{Fig:f500simulation}
\end{minipage}
\end{figure}

\begin{figure}
\begin{minipage}[t]{0.45\columnwidth} \centering
\includegraphics[width=0.9\columnwidth]{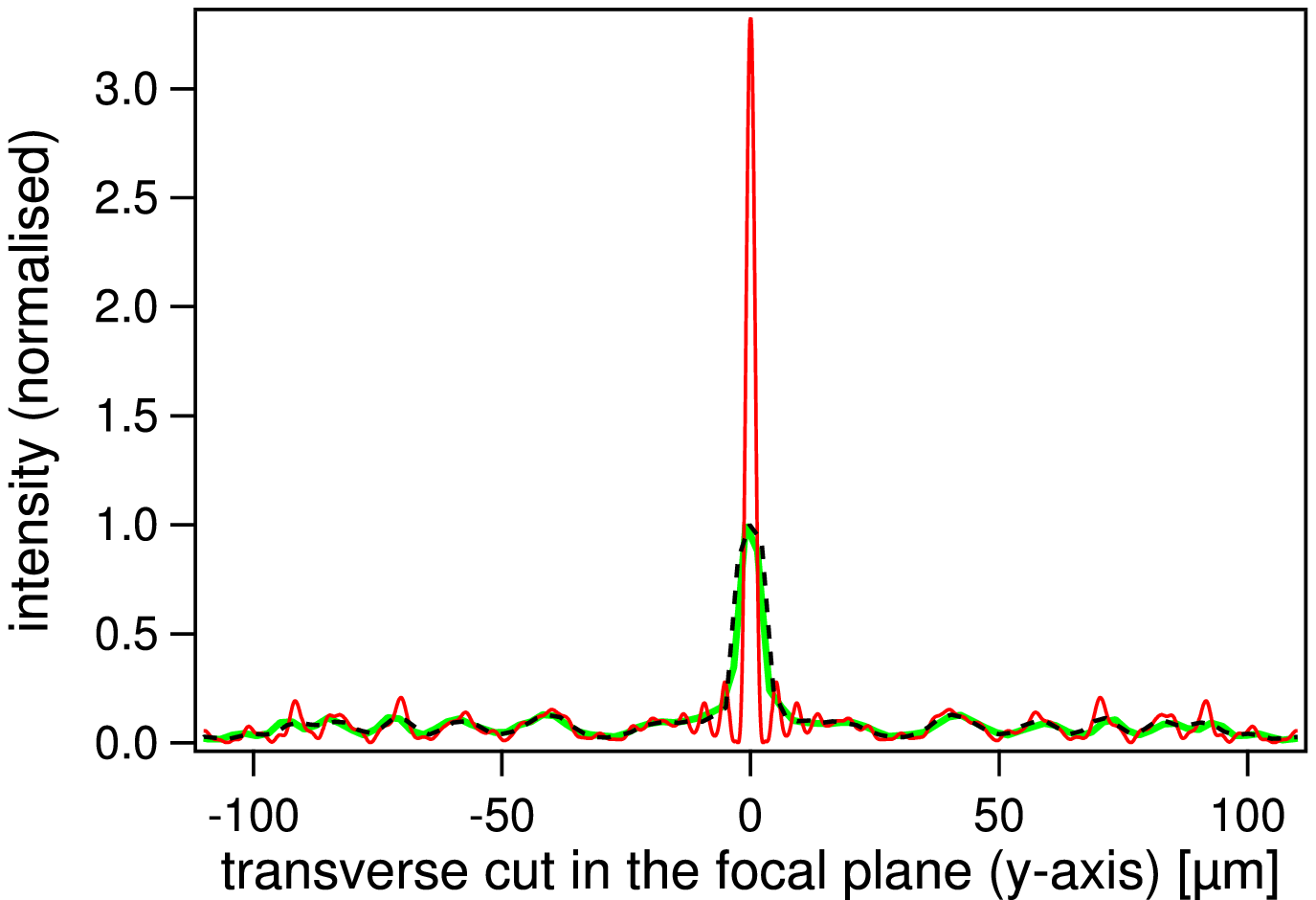}\caption{Intensity profiles in the
transverse focal plane ($y$-axis). Red curve shows the results of
the simulation in Fig.\,\ref{Fig:f500simulation}.  Black dashed
curve is the simulation averaged over the spatial resolution of
the optical measurement system.  Green curve is the measured
intensity, normalized to unity at the
peak.}\label{Fig:simplanfocal}
\end{minipage}\hspace{0.04\columnwidth}
\begin{minipage}[t]{0.45\columnwidth}\centering
\includegraphics[width=0.9\columnwidth]{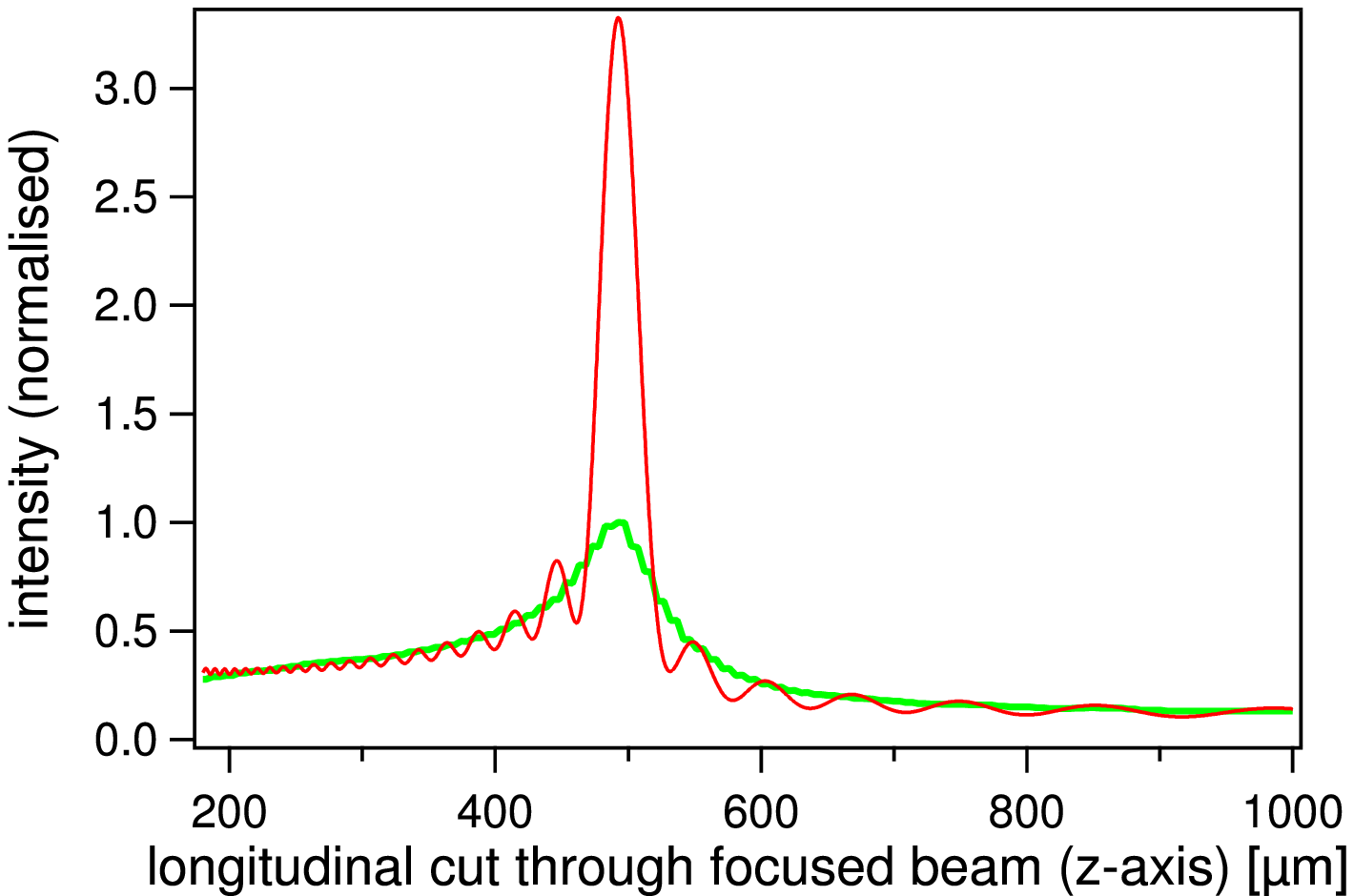}\caption{Intensity profile along the longitudinal
symmetry axis ($z$-axis).  Red curve shows the numerical
simulation of Fig.\,\ref{Fig:f500simulation}.  Green curve shows
the measured intensity, normalized to unity at the peak.
Dimensions of the simulated focus spot (FWHM) in the $x,y,z$
directions are are 192\,$\mu$m, 2.0\,$\mu$m, 35\,$\mu$m,
respectively.}\label{Fig:simaxeoptique}
\end{minipage}
\end{figure}

In order to verify the planar Fresnel lens designs we developed a
separate test setup that directly measures the focal properties of
these devices.  Figure \ref{Fig:imaging schematic} shows the
arrangement.
\begin{figure}
\includegraphics[width=0.75\columnwidth]{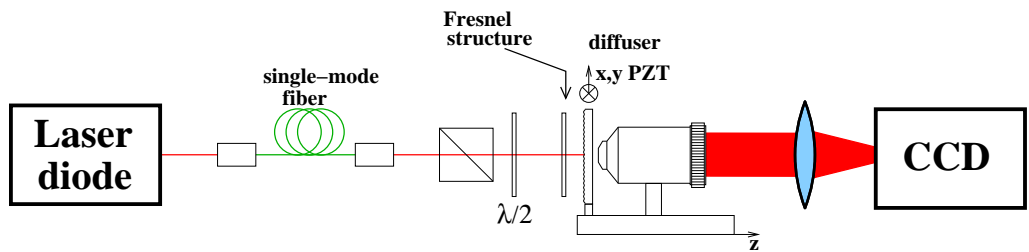}\caption{Schematic
of the Fresnel structure optical characterization system. The
piezoelectric element dithers the diffusing screen to eliminate
speckle in the spatial intensity distribution.}\label{Fig:imaging
schematic}
\end{figure}
A laser beam issuing from a stabilized laser diode coupled to a
single-mode fiber is linearly polarized and impinges on the
FIB-milled Fresnel structure from the back (substrate side). A
diffuser plate placed on the output side maps the spatial
intensity of the diffracted light in planes parallel to the
structure plane.  A piezoelectric element dithers the diffuser
plate in the $x-y$ plane to eliminate the effects of speckle, and
a microscope objective images the diffuser intensity pattern onto
a CCD camera. The distance between the
diffuser-plate-imaging-objective combination and the Fresnel
structure is systematically increased along $z$. At each distance
an image of the pattern is recorded on the CCD camera and
integrated along $x$, the long axis of the Fresnel slits, thereby
generating a series of profiles of the intensity distribution
along $y$ as a function of $z$. The resulting measured intensity
map in the $y-z$ plane is then compared with numerical simulations
to verify design accuracy. Figures \ref{Fig:f500experience} and
\ref{Fig:f500simulation} show the measured profile and the
numerical simulation respectively for the 500~$\mu$m focal length
Fresnel motif.
The intensity profiles in the transverse focal plane and along the
longitudinal axis are plotted in Figs.\,\ref{Fig:simplanfocal} and
\ref{Fig:simaxeoptique}.  These plots show that the simulated
transverse focus profile (red curve) is significantly sharper than
the measured CCD image (green curve).  The black-dashed curve
shows the simulated focal profile averaged over the effective
spatial resolution of 2.4~$\mu$m ($\times 10$ image magnification;
CCD pixel element 24~$\mu$m square). Overall dimensions of the
simulated focal spot along the $x,y,z$ directions are 192~$\mu$m,
2~$\mu$m, 35~$\mu$m, respectively.  The excellent agreement
between the measured profile and the spatially averaged calculated
profile indicates that the red-curve simulation well represents
the actual focus characteristic of the Fresnel lens. These results
support and encourage the use of numerical simulations to explore
more elaborate lens designs.
\subsection{Atom Trap Imaging System}
\begin{figure}
\includegraphics[width=0.50\columnwidth]{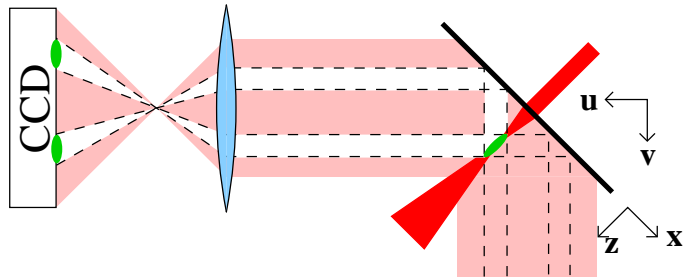}\caption{Schematic
of the absorption imaging system used to detect cold atoms in the
MMOT and in the FORT.  Green spot at focus of intense
off-resonance red laser beam represents trapped atoms.  Two green
spots on the CCD plane represent the absorption images focused by
the blue lens.  Note that  CCD plane is rotated by an angle of 45
 degrees with respect to the mirror plane.  Therefore trap images are foreshortened by $\sqrt{2}$ in the $z$ direction.}\label{Fig:detectionimagerie}
\end{figure}

\begin{figure}[h]
\begin{minipage}[t]{0.45\columnwidth} \centering
\includegraphics[width=0.95\columnwidth]{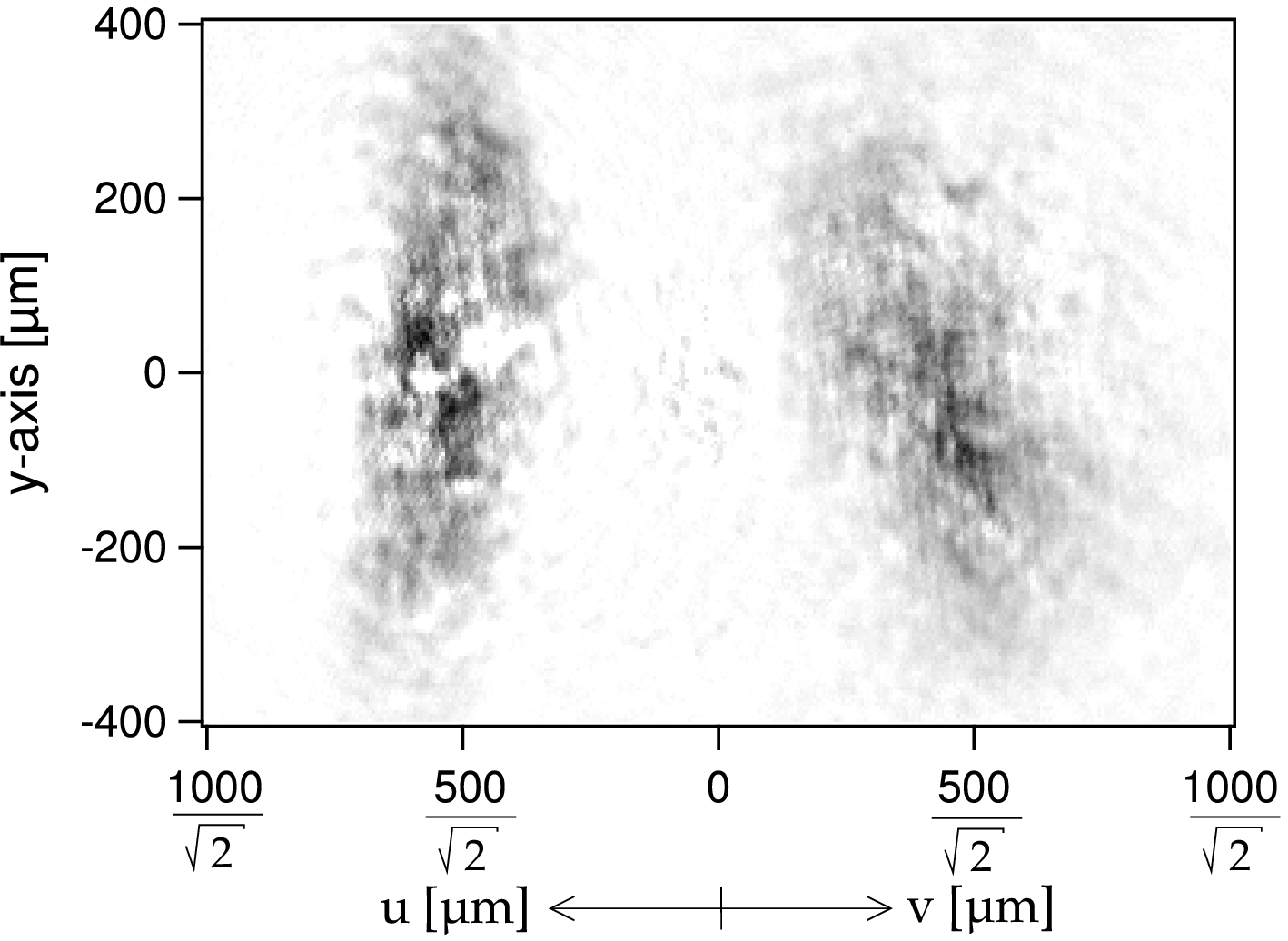}\caption{Double absorption image (see Fig.\,\ref{Fig:detectionimagerie}).
Left (right) image side is the projection of atom cloud on the
u-axis (v-axis), corrected by a $\sqrt{2}$ foreshortening.}
\label{Fig:MMOTimage}
\end{minipage}\hspace{0.04\columnwidth}
\begin{minipage}
[t]{0.45\columnwidth}\centering
\includegraphics[width=0.9\columnwidth]{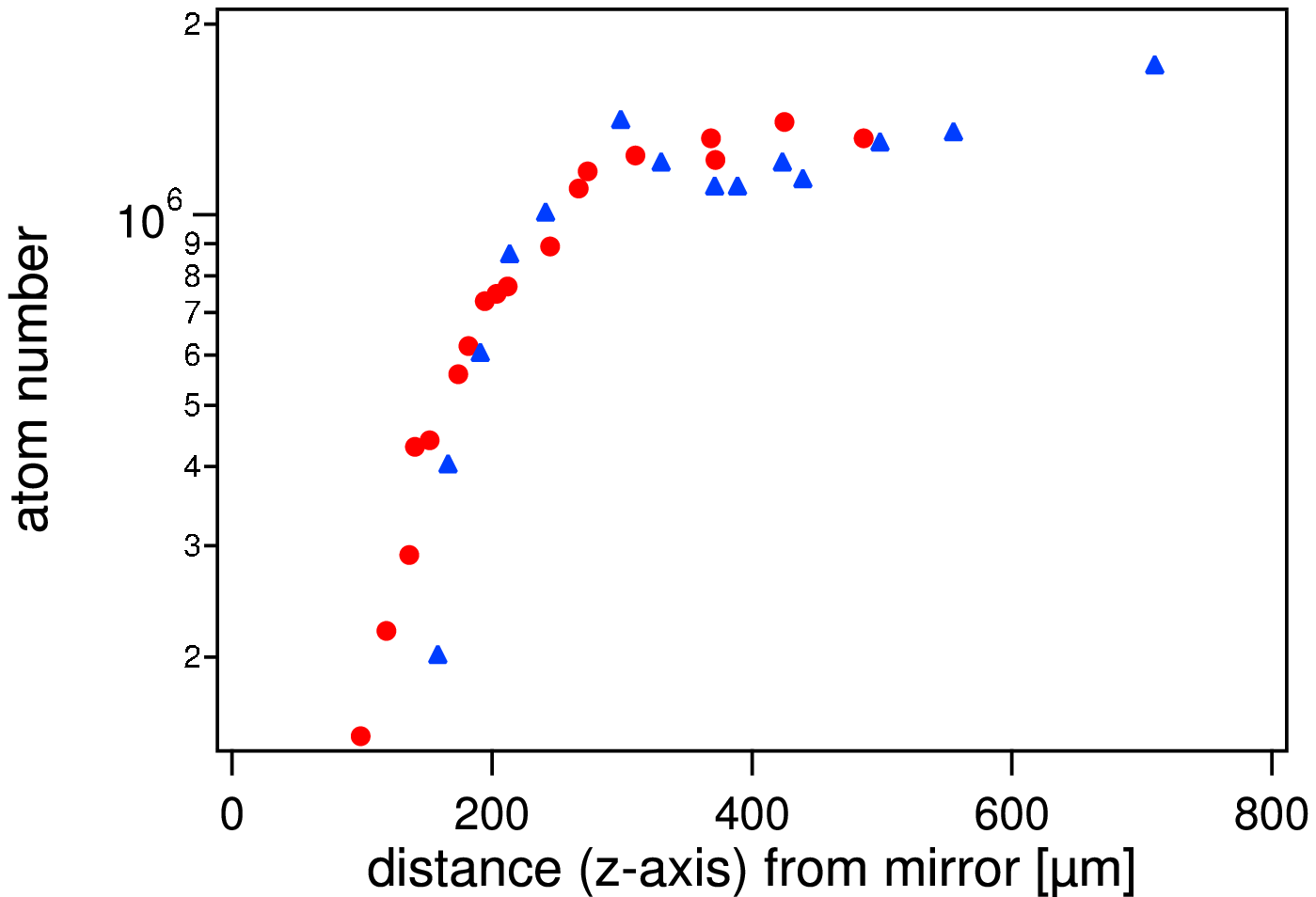}\caption{Number
of atoms contained in the MMOT as function of distance from the
mirror surface.  Blue triangles correspond to the mechanical
approach (via a vernier screw translation) of the mirror surface
to the MMOT, and red circles correspond to a translation of the
magnetic field minimum (via a small external bias
field).}\label{Fig:numberofatomsinMMOT}
\end{minipage}
\end{figure}

The atoms trapped in the MMOT and in the FFORT are imaged by an
absorption profile of a probe laser beam tuned near the
$\mathrm{F=4}\rightarrow \mathrm{F^{\prime}=5}$ transition. Figure
\ref{Fig:detectionimagerie} shows a schematic of the absorption
imaging system which registers a double image: one from atom
absorption followed by mirror reflection, the other from mirror
reflection followed by atom absorption.  We therefore obtain two
projections from orthogonal directions and thus a 3D view of the
atom distribution.
Figure \ref{Fig:MMOTimage} shows a typical image of the
distribution of cold atoms trapped in the MMOT when the trap
center is located $\simeq 500~\mu$m away from the mirror surface.
The optical thickness of  trapped atoms in the MMOT would be
sufficient to attenuate the probe laser beam almost to extinction
if tuned to the absorption resonance peak.  Therefore, for the
MMOT images, the probe is tuned off-resonance by about four
natural atomic line widths in order to maintain linearity between
absorption probability and atom number. For atoms trapped in the
FFORT, the imaging laser is resonant with the
$\mathrm{F=4}\rightarrow \mathrm{F^{\prime}=5}$ transition.  The
distance from the MMOT to the mirror plane can be varied either by
mechanically moving the mirror closer to the atoms with a vernier
screw adjustment or by adding a bias magnetic field that
translates the cloud center. Figure \ref{Fig:numberofatomsinMMOT}
shows that the minimum distance before atom loss from the trap
becomes significant is $\simeq 300 ~\mu$m.  We believe that the
onset of this loss occurs because the vertical beams of the MMOT
are partially occulted by the mirror.  At the center of the mirror
the resulting sharp edge diffraction produces a shadow about 100
$\mu$m wide that begins to perturb the MMOT loading rate at a
comparable distance from the mirror surface.

\label{imaging_pb}
Note that since the Fresnel lens is engraved on the mirror, all beams reflecting on the surface are diffracted by the structure. This includes the imaging beam, whose diffracted part is not collected by the imaging lens. 
There is therefore a region where we lose all information about the number of atoms. 
From geometrical considerations, the size of this blind region is half the width of the structure in the $x$ direction, so typically $\sim$~100~$\mu$m.
In the next generation of structures, this size will be reduced to obtain a 75~$\mu$m~x~200~$\mu$m footprint.
Absorption is thus decreased by a factor of 3 but this reduction permits imaging atoms down to 35-40 $\mu$m above the surface.

\section{FFORT atom loading from MMOT}
\subsection{Trapped atom number}
\begin{figure}
\begin{minipage}[t]{0.45\columnwidth}\centering
\includegraphics[width=0.9\columnwidth]{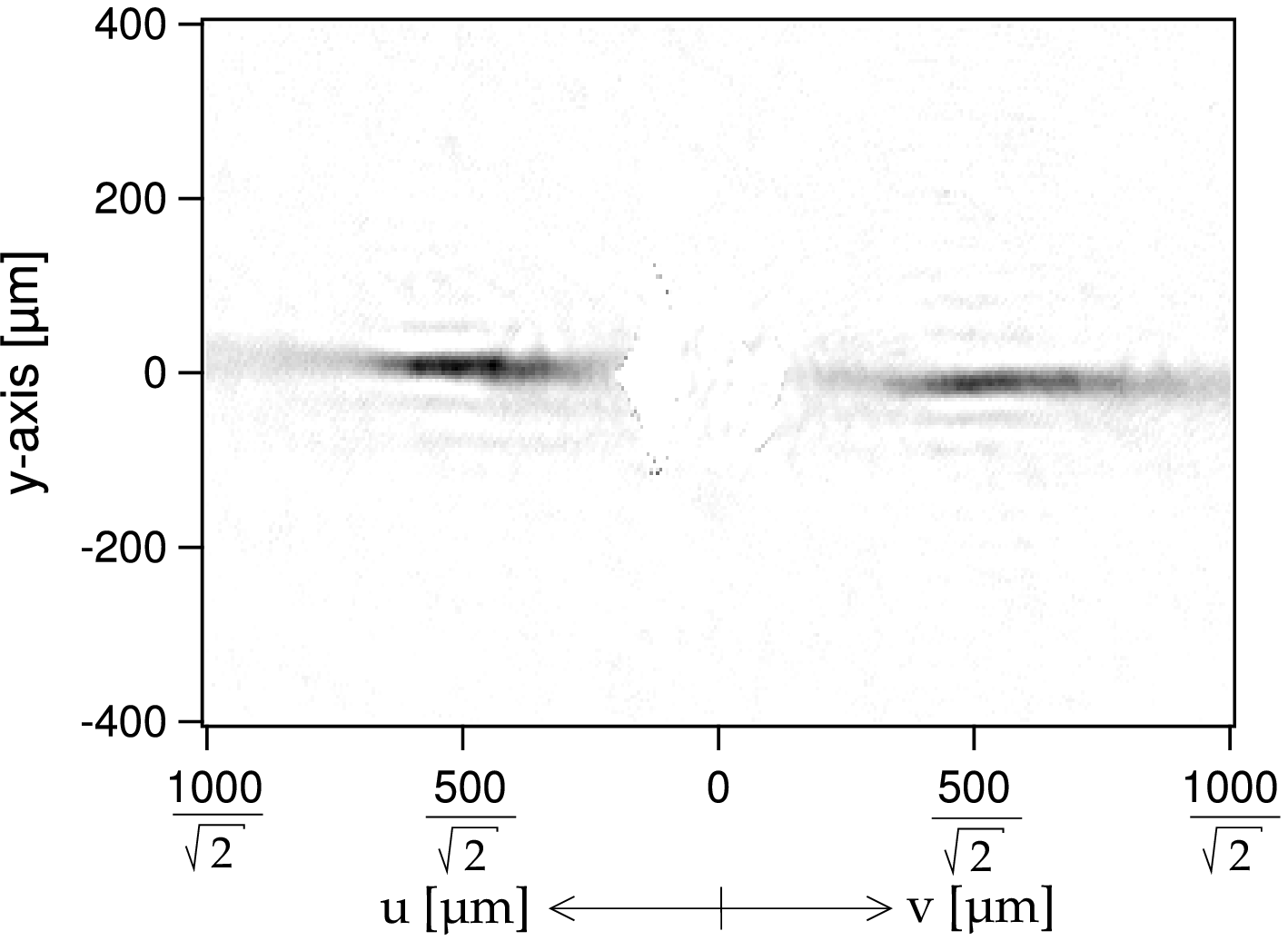}\caption{Double absorption image (see Fig.\,\ref{Fig:detectionimagerie}). Left(right) image side is the
projection of atom cloud on the u-axis (v-axis), corrected by a
$\sqrt{2}$ foreshortening.  Image recorded 30~ms after MMOT
extinction.}\label{Fig:FORTImage}
\end{minipage}\hspace{0.04\columnwidth}
\begin{minipage}[t]{0.45\columnwidth}\centering
\includegraphics[width=0.9\columnwidth]{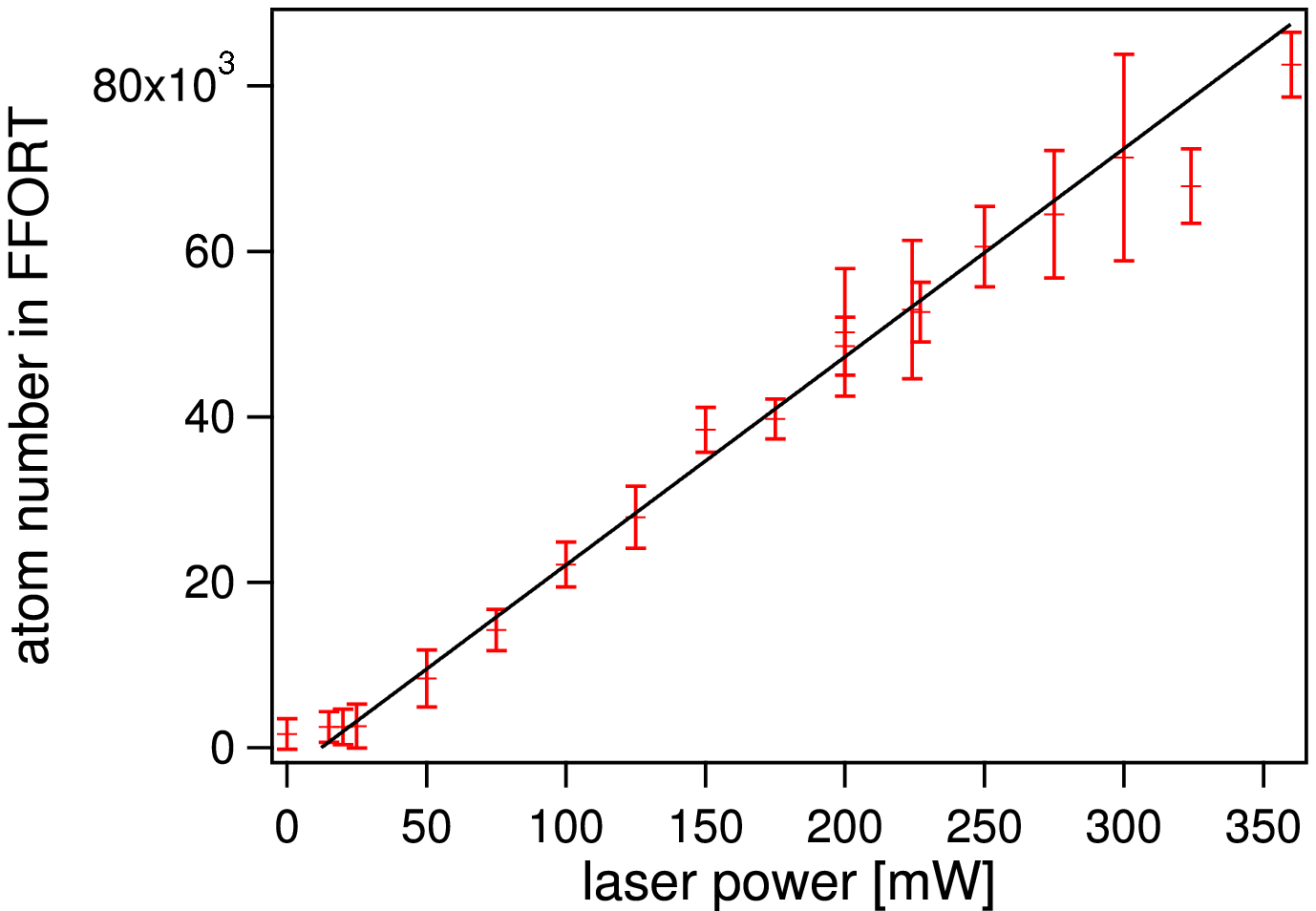}\caption{Atom
number vs. FFORT laser power.}\label{Fig:TrapPower}
\end{minipage}
\end{figure}

We have carried out a series of measurements to optimize FFORT
loading from the MMOT.  Figure \ref{Fig:FORTImage} shows an
absorption image of the atoms trapped in the FFORT with 500~$\mu$m
focal length.  We have also investigated 200~$\mu$m focal length
structures; but, in addition to atom loss from mirror proximity
(Fig.\,\ref{Fig:numberofatomsinMMOT}) technical limitations
associated with imaging discussed in section~\ref{imaging_pb}, restrict
systematic studies reported here to 500~$\mu$m focal-length
devices.

We first measured the trapped atom number as a function of FFORT
laser power. The results are shown in Fig.\,\ref{Fig:TrapPower}
and, above a threshold of 20~mW, indicate a linearly increasing
number of atoms.  As the trapping laser power grows, the effective
volume of the FFORT grows as well --- first in the central focal
region, then in the small secondary maxima adjacent to the main
peak (see
Figs.\,\ref{Fig:simplanfocal},\,\ref{Fig:simaxeoptique}). For
these measurements the dipole-gradient trap laser detuning was
fixed at $\simeq 0.5$~nm to the red of the Cs resonance line.

Next Fig.\,\ref{Fig:TrapDetuning} plots the atom number in the
FFORT as a function of red detuning.  We found that a detuning of
$\simeq -0.25$~nm produced the optimum atom number.  The rapid
fall-off at smaller detunings is due to absorptive heating while
at larger detunings it is primarily due to the inverse relation
between trap depth and detuning. The data shown in
Figs.\,\ref{Fig:TrapPower},\,\ref{Fig:TrapDetuning} were recorded
30~ms after extinguishing the MMOT so as to purge untrapped atoms
from the imaging field of view.

\subsection{Loading efficiency}
We also determined loading efficiency, defined as the ratio of
trapped atoms to the initial atom number in the MMOT.  The
relatively poor spatial overlap between the MMOT and the FFORT
(compare Figs.\,\ref{Fig:MMOTimage},\,\ref{Fig:FORTImage}),
results in a rather low loading efficiency of only a few per cent.

In order to evaluate the number of trapped atoms initially in the volume of the diffraction pattern of the lens, we carried out an optical pumping experiment through the structure.
With the dipole trap laser off, we first pump all MMOT atoms (Cs~[$\mathrm{S}_{1/2},\mathrm{F}$=4]) to the $\mathrm{F}$=3 state with a 200~$\mu$s pulse resonant on the 
$\mathrm{F}=4 \rightarrow \mathrm{F^{\prime}}=4$ transition.
Then, we illuminate the structure from behind with a 
$\mathrm{F}=3 \rightarrow \mathrm{F^{\prime}}=3$ resonant 200~$\mu$s pulse, which pumps the atoms back to the $\mathrm{F}$=4 state.
As this laser goes through the lens, only the atoms in the diffraction pattern are affected.
We then take a usual absorption image on the $\mathrm{F}=4 \rightarrow \mathrm{F^{\prime}}=5$ transition.
Atoms outside the diffraction pattern are still in the $\mathrm{F}=3$ state, and thus transparent to this probe.
Comparing the number of atoms in the central peak of this distribution to the number of atom in the FFORT, we can conclude that the capture efficiency of the dipole trap approaches unity.

\subsection{Temperature after release}
\begin{figure}
\begin{minipage}[t]{0.45\columnwidth}\centering
\includegraphics[width=0.9\columnwidth]{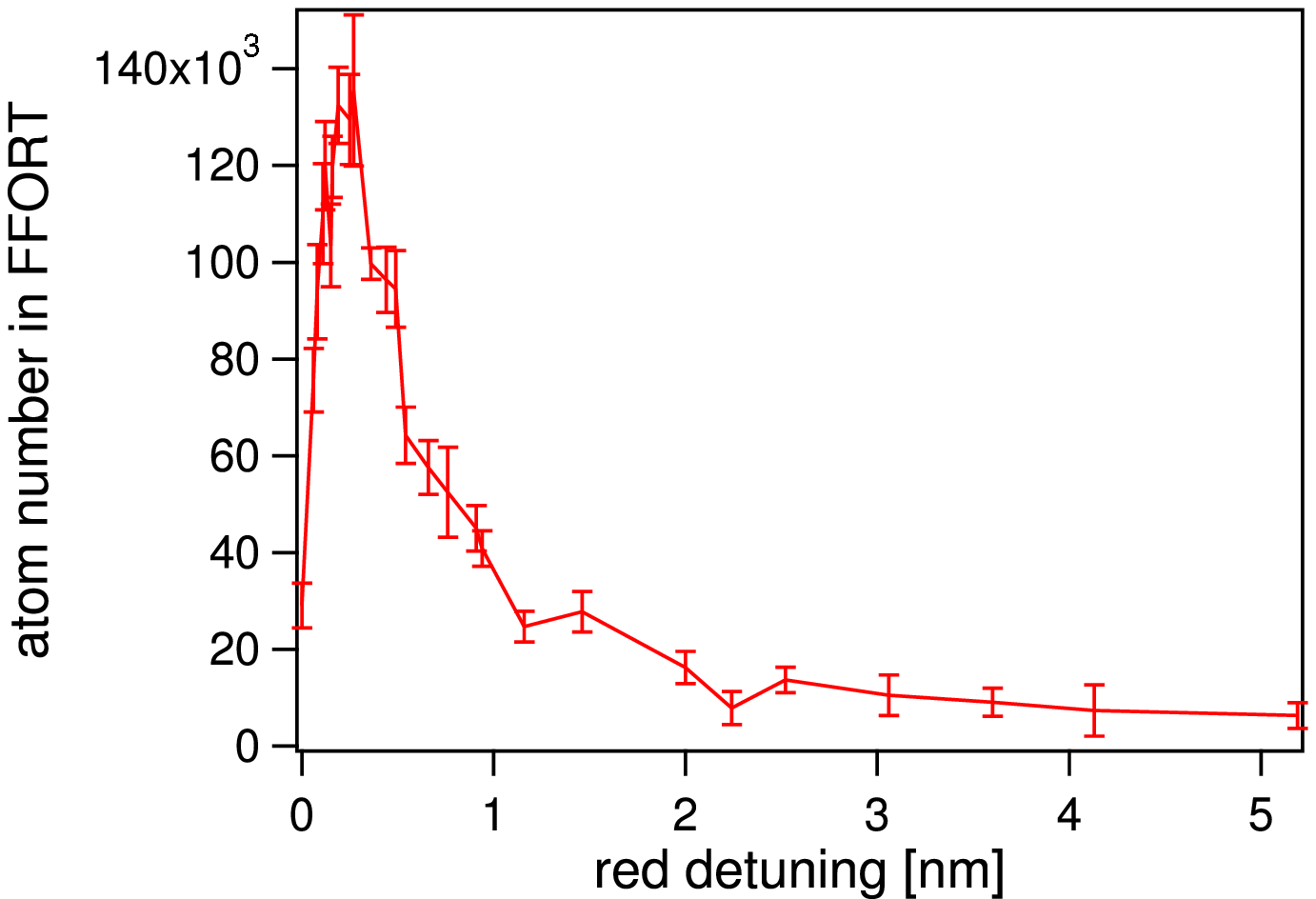}\caption{Atom number vs. FFORT detuning.}
\label{Fig:TrapDetuning}
\end{minipage}\hspace{0.04\columnwidth}
\begin{minipage}[t]{0.45\columnwidth}\centering
\includegraphics[width=0.9\columnwidth]{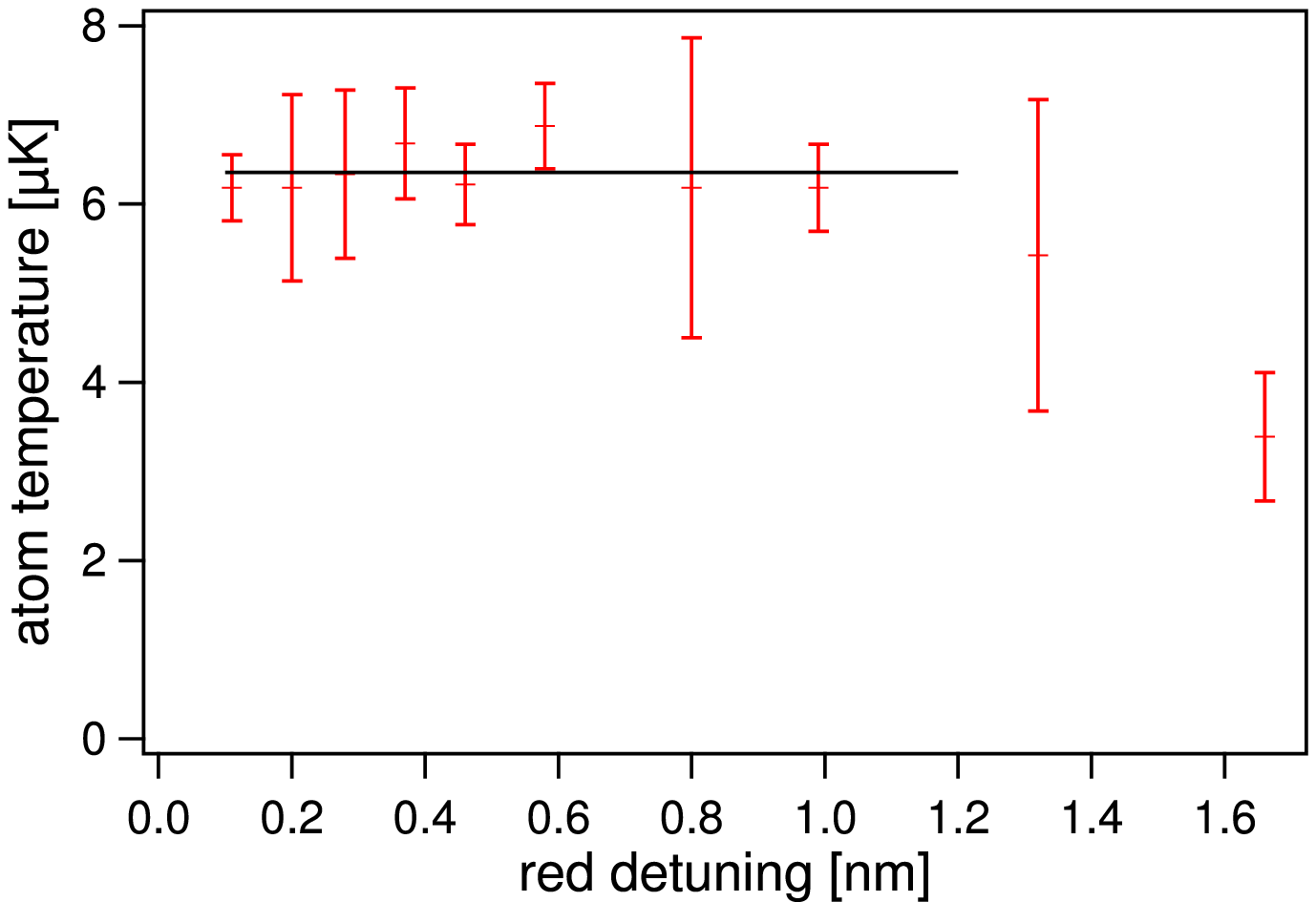}\caption{Trap
temperatures vs. FFORT red detuning.}\label{Fig:TempTrapDetuning}
\end{minipage}
\end{figure}

Finally we also measured the atom temperature in the confining $y$
direction within the trap by monitoring the atom cloud ballistic
expansion after trap release\,\cite{GW00}. The results, as a
function of trap detuning are shown in
Fig.\,\ref{Fig:TempTrapDetuning}.  The FFORT-trapped atoms are
found to be significantly colder than the atoms in the MMOT ($\sim
6~\mu$K vs. $\sim 30~\mu$K). The origin of this extra cooling
effect has not yet been thoroughly investigated.  However from
numerical simulation we estimate the trapping frequency along the
confining axis to be about 13~kHz.  Since a mechanical shutter
extinguishes the FFORT in about 1~ms, we speculate that part of
the cooling effect comes from adiabatic expansion just prior to
ballistic release. Note also that the temperature starts to
decline with detuning beyond $\sim$ 1~nm, where decreasing trap
depth might lead to some evaporative cooling effect.
\section{Perspectives}
\begin{figure}
\begin{minipage}[t]{0.45\columnwidth}\centering
\includegraphics[width=0.9\columnwidth]{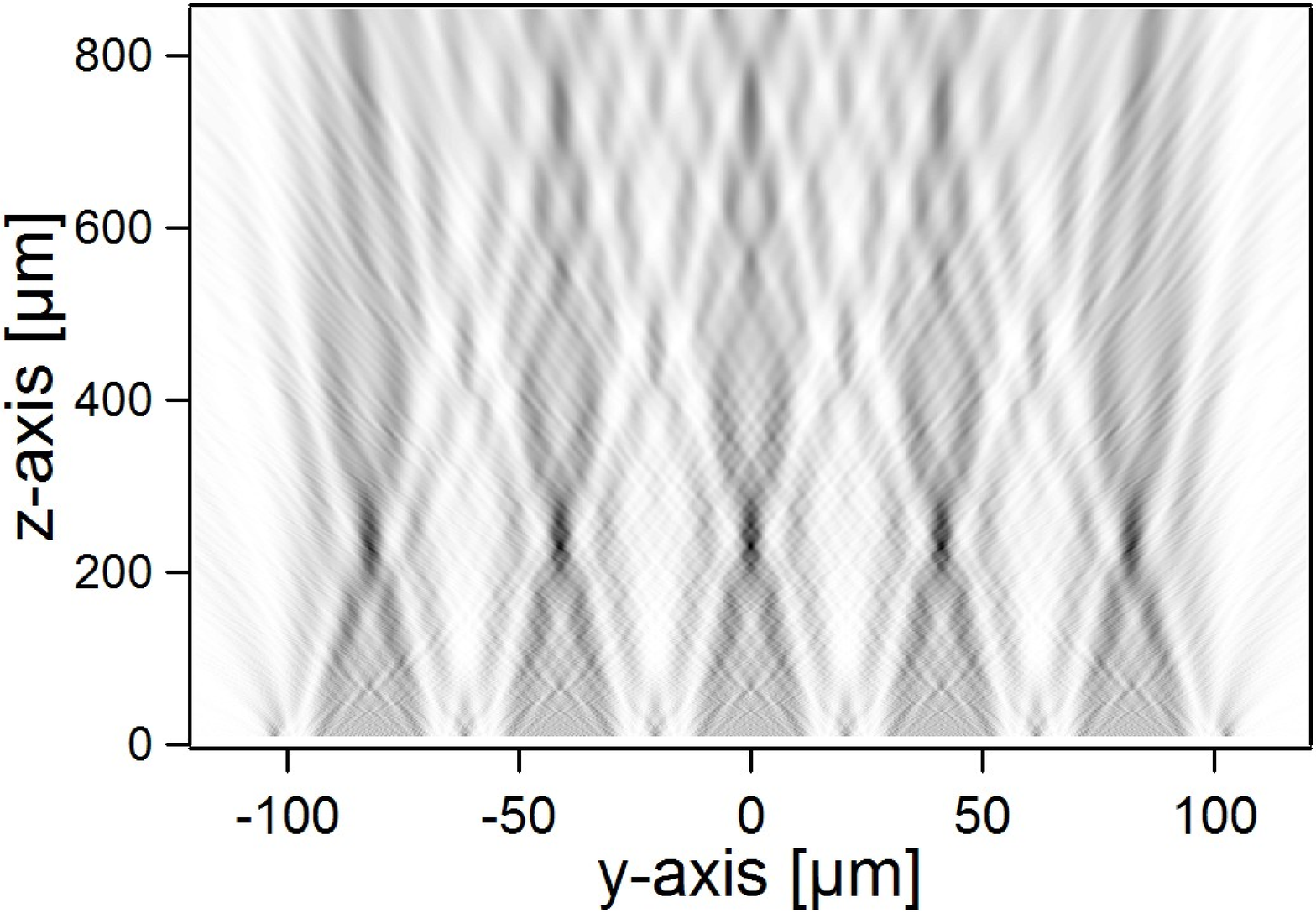}\caption{Numerical simulation of the diffractive
pattern for a 1-D array of five 200~$\mu$m focal length structures
measured in
Fig.\,\ref{Fig:5motif-array-meas}.}\label{Fig:5motif-array-cart}
\end{minipage}\hspace{0.04\columnwidth}
\begin{minipage}[t]{0.45\columnwidth}\centering
\includegraphics[width=0.9\columnwidth]{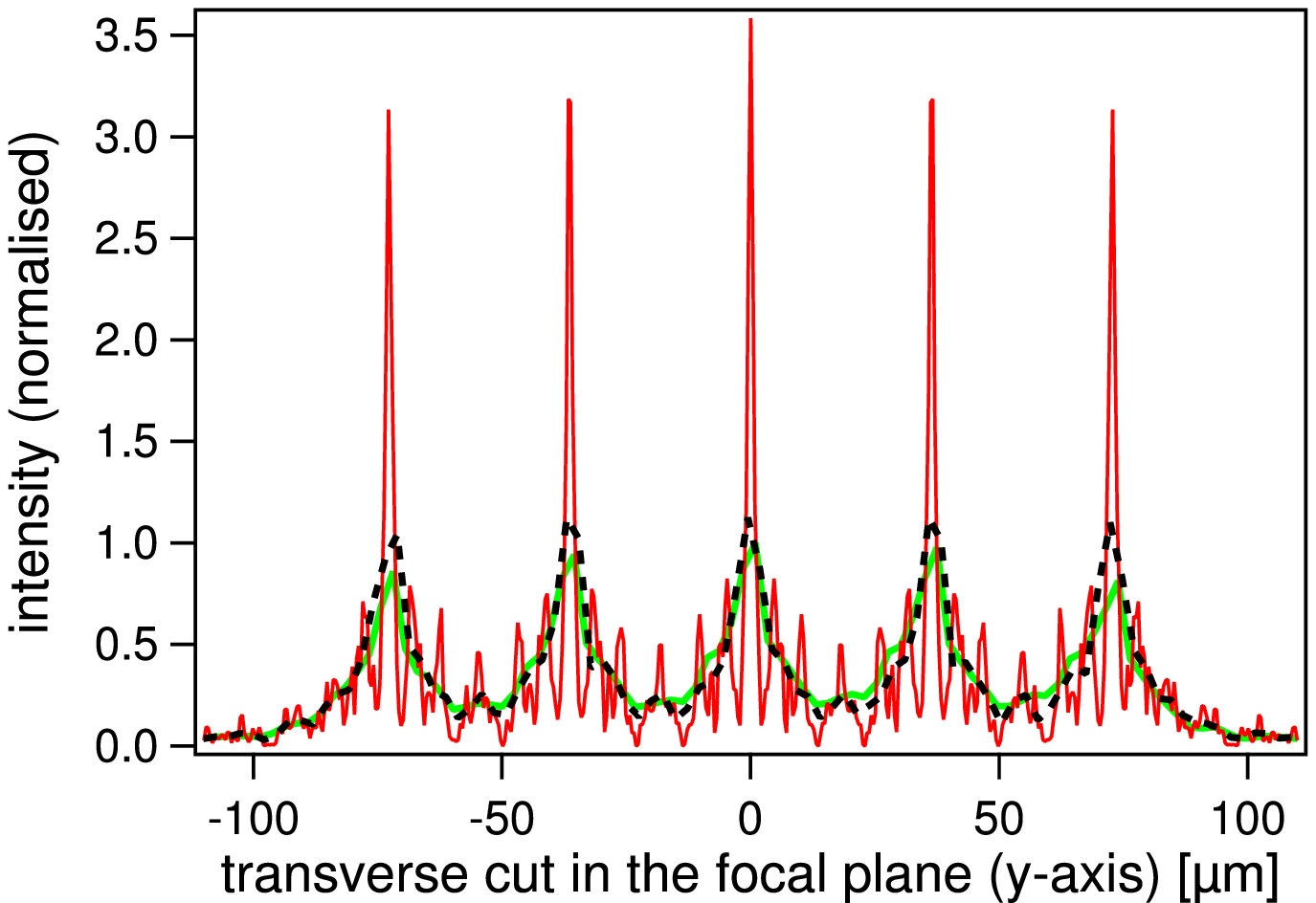}\caption{Intensity profile in the
transverse focal plane ($y$-axis) for 1D array of five 200 $\mu$m
focal length lenses. Red curve shows the results of the simulation
in Fig.\,\ref{Fig:5motif-array-cart}. Black dashed curve is the
simulation averaged over the spatial resolution of the optical
measurement system.  Green curve is the measured intensity,
normalized to unity at the peak.}\label{Fig:5motif-array-meas}
\end{minipage}
\end{figure}
Several different directions for development of these planar,
all-optical atom trapping structures remain to be exploited. As
the work presented here shows, the most promising involve not only
miniaturisation but \emph{integration} of structure functionality
directly onto the chip.  First, we are investigating 1D trap
arrays as shown in Figs.\,\ref{Fig:5motif-array-cart} and
\ref{Fig:5motif-array-meas} and 2D arrays as well.  We are also
developing variable-focal-length lenses useful for capturing atoms
far from the mirror and guiding them into very small volumes close
to the surface.  Second, we are exploring the integration of the
optical structures with on-chip microelectronic circuitry and
micromechanical (MEMS) devices for each array element. Dynamic
addressing of the trapped cold atoms either electrically with
integrated current-carrying wires or optically with laser spots
provide typical examples.  Finally, miniaturization can be greatly
improved by reducing the size of FIB-milled Fresnel elements while
maintaining adequate trapping efficiency. Further reduction to
about a 10~$\mu$m footprint should be possible by taking advantage
of subwavelength surface wave phenomena\,\cite{LDD02,GAV06}.
Development of planar arrays together with atom transport and
dynamic array-element addressing, opens the way to applications in
quantum gate implementation and precision atom doping of surfaces.

\begin{acknowledgments}
Support from the Minist{\`e}re d{\'e}l{\'e}gu{\'e} {\`a} l'Enseignement sup{\'e}rieur et {\`a}
la Recherche under the programme
ACI-``Nanosciences-Nanotechnologies," the R{\'e}gion Midi-Pyr{\'e}n{\'e}es
[SFC/CR 02/22], and FASTNet [HPRN-CT-2002-00304]\,EU Research
Training Network, is gratefully acknowledged.  Technical
assistance of H. Lezec and the fabrication facilities of the
Caltech Kavli Nanoscience Institute are also gratefully
acknowledged.
\end{acknowledgments}

\end{document}